\documentclass[twocolumn,showpacs,preprintnumbers,amsmath,amssymb]{revtex4}
\usepackage{graphicx}% Include figure files
\usepackage{dcolumn}% Align table columns on decimal point
\usepackage{bm}% bold math

\begin{document}

\preprint{Submit to APL}

\title{Electrical coherent control of nuclear spins in a breakdown regime of quantum Hall effect}% Force line breaks with \\

\author{H. Takahashi, M. Kawamura, S. Masubuchi, K. Hamaya, and T. Machida$\footnote{Electronic mail: tmachida@iis.u-tokyo.ac.jp }$$\footnote{Also at INQIE, University of Tokyo.}$}%
\affiliation{Institute of Industrial Science, University of Tokyo,  4-6-1 Komaba, Meguro-ku, Tokyo 153-8505, Japan}

\author{Y. Hashimoto and S. Katsumoto$^{\dagger}$}
\affiliation{Institute for Solid State Physics, University of Tokyo, 5-1-5 Kashiwanoha, Kashiwa, Chiba 277-8581, Japan}

\date{\today}% It is always \today, today,
\begin{abstract}
Using a conventional Hall-bar geometry with a micrometal strip on top of the surface, the authors demonstrate an electrical coherent control of nuclear spins in an AlGaAs/GaAs semiconductor heterostructure. A breakdown of integer quantum Hall (QH) effect is utilized to dynamically polarize nuclear spins. By applying a pulse rf magnetic field with the metal strip, the quantum state of the nuclear spins shows Rabi oscillations, which is detected by measuring longitudinal voltage of the QH conductor.
\end{abstract}
\pacs{Valid PACS appear here}% PACS, the Physics and Astronomy
\maketitle
For quantum information processing, it is essential to control quantum states without destroying their coherence. Since nuclear spins have extremely long coherence times arising from weak couplings with their environments, application of the nuclear spins as quantum bits (qubits) has been discussed.\cite{Kane,Ladd} In fact, Vandersypen {\it et al}.\cite{Vandersypen} demonstrated seven-qubit quantum computation using nuclear magnetic resonance (NMR) of liquid-state molecules. In a solid, though it is difficult to individually access and manipulate a single nuclear spin in practice, a theoretical study suggests that ensemble of nuclear spins can be utilized collectively as a long-lived quantum memory for qubits.\cite{Taylor} 

In AlGaAs/GaAs semiconductor devices, the ensemble of nuclear spins is often polarized by flip-flop processes of electron and nuclear spins via the hyperfine interaction.\cite{Salis,MachidaIE,Yusa,Hashimoto,Smet} By combining a pulse rf NMR technique with dynamic nuclear polarization (DNP) along spin-resolved quantum Hall (QH) edge channels, we have achieved electrical coherent manipulation of their quantum states.\cite{MachidaIE} Recently, Yusa {\it et al.}\cite{Yusa} also demonstrated coherent manipulation of nuclear-spin ensemble using a fractional QH bulk region of Landau-level filling factor $\nu =$ 2/3, where the electron system is separated into spin-polarized and spin-unpolarized domain structures due to competition between Zeeman energy and Coulomb exchange interaction.\cite{Hashimoto,Smet} Their study reported multiple quantum coherence of nuclear spins.

In this letter, in order to dynamically polarize nuclear spins, we utilize a breakdown regime of integer QH effect of $\nu =$ 1, in which the mechanism of the DNP is clear.\cite{Kawamura} Using a conventional Hall-bar sample made of an AlGaAs/GaAs semiconductor, we demonstrate coherent manipulation of the quantum state of nuclear spins; by combining a pulse rf NMR technique with the DNP, we observe evident Rabi oscillations.
\begin{figure}[b]
\includegraphics[width=8.5cm]{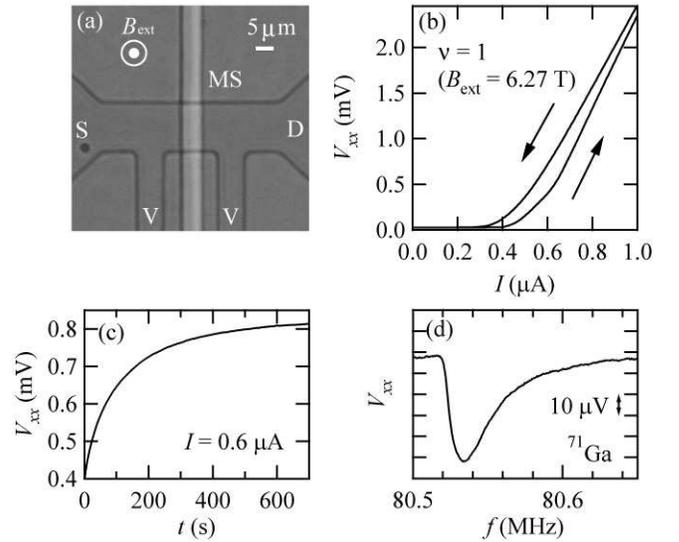}
\caption{(a) Optical micrograph of a central region of the device. (b) $V_{xx}$ vs $I$ at $\nu$ = 1 ($B_\text{ext}$ = 6.27 T) for the sweep directions depicted by arrows. (c) Time evolution of $V_{xx}$ at $I$ = 0.6 $\mu$A. (d) A NMR spectrum for $^{71}$Ga detected by measuring $V_{xx}$ at $I$ = 0.6 $\mu$A.}
\end{figure} 

A Hall-bar device we studied here was fabricated on an Al$_{0.35}$Ga$_{0.65}$As/GaAs single heterostructure with an electron mobility of 220\,m$^{2}$/V s and a sheet electron density of $1.6\times10^{15}$\,m$^{-2}$. Figure 1(a) shows an optical micrograph of a central region of the device. The two voltage probes (V) were located between the source and drain reservoirs (S and D). A 5-\,$\mu$m-wide metal strip (MS) was prepared by depositing 100-nm-thick Au and 5-nm-thick Ti on top of the surface. Nuclear spins were manipulated by applying rf current to the MS, which generates rf magnetic fields $B_\text{rf}$. All measurements were performed by a standard dc four-terminal method in a $^3$He-$^4$He dilution refrigerator. The measurements for Figs. 1-3 were carried out at 50\,mK.  

To initialize nuclear spins, we make use of a method for electrical DNP, which was developed in our recent work.\cite{Kawamura} When the current ($I$) flowing a QH device is increased above a critical current, electrons are excited to an upper Landau level, giving rise to an abrupt increase in longitudinal voltage $V_{xx}$. This phenomenon is referred to as QH effect breakdown. In a breakdown regime of odd-integer QH effect, the electron excitation involves up-to-down spin-flip processes. As a consequence, nuclear spins can be dynamically polarized (down to up) over the bulk region of the QH device through the hyperfine interaction.\cite{Kawamura} The polarized nuclear spins reduce the spin-splitting energy of $\Delta E=|g|\mu_\text{B}B$, where $g$ is the g factor of electrons (= -0.44 in GaAs) and $\mu_\text{B}$ is the Bohr magneton. Hence, this process can accelerate the QH effect breakdown, leading to an increase in the $V_{xx}$ and hysteretic $I$-$V_{xx}$ characteristics depending on the sweep direction.\cite{Kawamura} 
\begin{figure}[t]
\includegraphics[width=7cm]{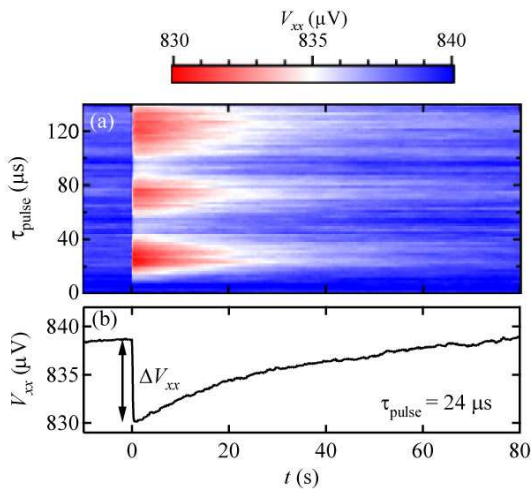}
\caption{(Color online) (a) Color scale plot of $V_{xx}$ as functions of rf pulse duration $\tau_\text{pulse}$ and measurement time $t$ for $V_\text{rf} =$ 4.75 V. (b) A representative time trace of $V_{xx}$ for $\tau_\text{pulse} =$ 24 \,$\mu$s, where the pulse is applied at $t$ = 0\,s.}
\end{figure}  

We apply an external magnetic field of $B_\text{ext}$= 6.27 T ($\nu = 1$) in order to use the above DNP method. Figure 1(b) displays $V_{xx}$ versus $I$, here $I$ was swept between -1 and 1\,$\mu$A at a rate of 17 nA/s. Clear increase in $V_{xx}$ with increasing $I$, i.e., QH effect breakdown,  is seen. Also, a pronounced hysteretic feature depending on the sweep direction is observed, being similar to the data in Ref.\cite{Kawamura}. When we applied $I = 0.6\,\mu$A to initialize nuclear-spin state, the increase in the $V_{xx}$ is saturated in several minutes [see Fig. 1(c)]. In this procedure, we now expect that the nuclear spins are polarized over the bulk region of the Hall-bar device. By applying continuous-wave $B_\text{rf}$, we can observe a NMR spectrum with a dip structure for $^{71}$Ga nuclei at a frequency of 80.529\,MHz, shown in Fig. 1(d). Since the amplitude of $B_\text{rf}$ perpendicular to the $B_\text{ext}$ is strong only in the region beneath the metal strip line, the nuclear spins beneath the metal strip can be controlled exclusively by the $B_\text{rf}$.\cite{Machida2} NMR spectra for $^{69}$Ga and $^{75}$As nuclei were also measured (not shown here).

In the following, we manipulate $^{71}$Ga nuclei using a pulse NMR method:\cite{MachidaIE,Yusa,Sanada} nuclear-spin state evolves with time during the application of $B_\text{rf}$ with pulse durations $\tau_\text{pulse}$. To measure the time evolution data, the rf pulses with different durations (0 $\leqslant$ $\tau_\text{pulse}$ $\leqslant$ 300\,$\mu$s) were applied at a peak-to-peak rf-voltage amplitude of $V_\text{rf}$. Figure 2(a) shows a color scale plot of $V_{xx}$ as functions of $\tau_\text{pulse}$ and measurement time ($t$) for $^{71}$Ga nuclei, where the pulses are irradiated at $t$ = 0\,s. After the irradiation, we can see abrupt decreases in the $V_{xx}$ (red regions) for various $\tau_\text{pulse}$. Representative time evolution data for $\tau_\text{pulse} =$ 24\,$\mu$s are shown in Fig. 2(b). The reduced $V_{xx}$ at $t$ = 0\,s is recovered at the time scale of several tens of seconds. As described in this figure, the sudden reduction in $V_{xx}$ is denoted by $\Delta V_{xx}$ for all the $\tau_\text{pulse}$.
\begin{figure}[t]
\includegraphics[width=8.5cm]{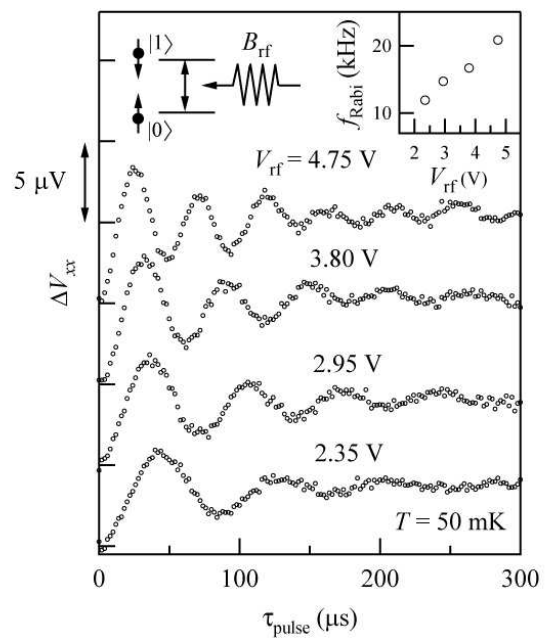}
\caption{Rabi oscillations traced by the amplitude of the $\Delta V_{xx}$ for various $V_\text{rf}$. The curves are offset for clarity. The inset shows $f_\text{Rabi}$ vs $V_\text{rf}$.}
\end{figure}
\begin{figure}[t]
\includegraphics[width=8.5cm]{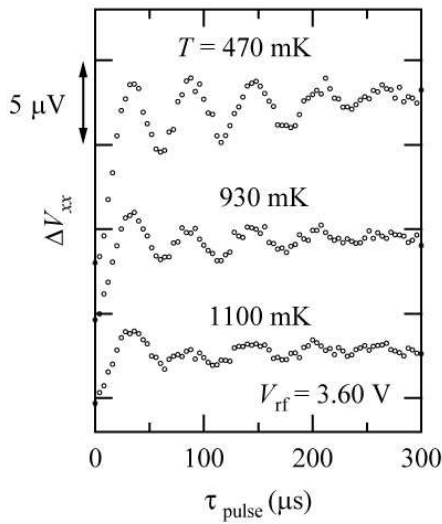}
\caption{Rabi oscillations traced by the amplitude of the $\Delta V_{xx}$ at various $T$. The curves are offset for clarity.}
\end{figure} 

The top plot of Fig. 3 shows $\Delta V_{xx}$ as a function of $\tau_\text{pulse}$ at $V_\text{rf} =$ 4.75 V [Fig. 2(a)]. A clear periodic oscillation is seen, and the oscillation amplitude is gradually decayed with a decay time of $\sim$ 100 $\mu$s. We further conduct similar measurements for various $V_\text{rf}$ and show the obtained results in this figure. We can see an increase in the periodic oscillation frequency ($f_\text{Rabi}$) with increasing $V_\text{rf}$, and the $f_\text{Rabi}$ is proportional to $V_\text{rf}$ as shown in the inset of Fig. 3. Therefore, these results described in Fig. 3 indicate coherent evolution of the quantum superposition in the time domain, i.e., Rabi oscillations. We also observe such Rabi oscillations at elevated temperatures, as shown in Fig. 4. Note that the $\Delta V_{xx}$ clearly oscillates at a temperature even above 1 K.

We infer that the decay time of the Rabi oscillations described above is restricted to the interaction between electron and nuclear spins. Recently, we achieved the Knight shift measurements of the NMR spectra in QH edge channels\cite{Masubuchi2} and bulk regions,\cite{Kawamura2} in which the hyperfine interaction between electron and nuclear spins is decoupled by gate-bias voltages during $B_\text{rf}$ application. This technique will enhance the lower limit of coherence time. For further enhancement, decoupling techniques of dipole-dipole interaction between nuclei should be taken into account.\cite{Yusa2,Sanada} 

We comment on advantages of this method for coherent control of nuclear-spin state in solids as follows. First, as shown in Fig. 4, this technique enables us to realize its demonstration at higher temperatures where the fractional QH effect cannot be observed. Second, the mechanism of DNP is clear: the DNP can be induced by just applying the current to the $\nu = 1$ QH system (other odd-integer QH systems can be also utilized\cite{Kawamura}), and the polarity of the polarized nuclear spins can be understood by the change in the $V_{xx}$. Namely, the mechanism of the resistance changes is clear. Third, a very conventional and macroscopic Hall-bar sample with a metal strip is merely needed. Last, it is not essential to get ultrahigh quality AlGaAs/GaAs heterostructures. Therefore, we can extend this method for QH conductors based on other materials.

In summary, we have demonstrated coherent control of nuclear spins using a conventional semiconductor Hall-bar device with a micrometal strip. A breakdown regime of integer QH effect combined with a pulse rf NMR technique is utilized to polarize and manipulate the nuclear spins. Clear Rabi oscillations are gained by measuring the change in the longitudinal voltage. 

This work is supported by the Grant-in-Aid from MEXT and the Special Coordination Funds for Promoting Science and Technology. One of the authors (K.H.) acknowledges JSPS Research Fellowships for Young Scientists.


%\begin{flushright}
%{\large Figure 1\\
%H. Takahashi\\
%Phys. Rev. B}%
%\end{flushright}
%\clearpage

%\begin{flushright}
%{\large Figure 2\\
%H. Takahashi\\
%Phys. Rev. B}%
%\end{flushright}

%\clearpage

%\begin{flushright}
%{\large Figure 3\\
%H. Takahashi\\
%Phys. Rev. B}%
%\end{flushright}

%\clearpage

%\begin{flushright}
%{\large Figure 4\\
%H. Takahashi\\
%Phys. Rev. B}%
%\end{flushright}

%\clearpage

%\begin{flushright}
%{\large Figure 5\\
%H. Takahashi\\
%Phys. Rev. B}%
%\end{flushright}

%\begin{flushright}
%{\large Figure 6\\
%H. Takahashi\\
%Phys. Rev. B}%
%\end{flushright}

%\newpage %Just because of unusual number of tables stacked at end
\bibliography{apssamp}% Produces the bibliography via BibTeX.

\end{document}